\documentclass{article}
\usepackage{graphicx}
\usepackage{newcent} 
\usepackage{fancyhdr}
\usepackage{hyperref} 
\usepackage{cite}
\usepackage{charter} 
\usepackage{multicol}
\usepackage{amsmath}
\usepackage{caption}
\usepackage{subcaption}

\usepackage[a4paper,top=2cm,bottom=2cm,left=3cm,right=3cm,footskip=.8cm,truedimen,headheight=17pt]{geometry}
\hypersetup{colorlinks=true,citecolor=red,urlcolor=blue}
\begin{document}
\title{Pauli crystals in harmonic trap and on a sphere}%
\author{Filip Gampel, Mariusz Gajda,  
\\
Magdalena Za{\l}uska--Kotur, Jan Mostowski \\[0.1cm]
           \small Institute of Physics of the Polish Academy of Sciences \\[-0.1cm]
           \small Al. Lotnik\'ow 32/46, 02-668 Warsaw, Poland}
\maketitle
\thispagestyle{fancy}

\begin{abstract}
Recently predicted \cite{2016GajdaEPL} and observed \cite{PhysRevA.97.063613, PhysRevLett.126.020401} Pauli crystals are structures formed by trapped ultracold non-interacting particles obeying Fermi statistics. The relative positions of the particles are determined by the trapping potential and the Pauli exclusion principle. Measuring all particles at once, their positions tend to be close to vertices of non-trivial polyhedrons. Similar shapes are found for systems of particles bound to the surface of a sphere. 
\end{abstract}

\section{Introduction}
In this paper we study a simple many-body problem. We consider a small group of non-interacting ultra-cold particles obeying Fermi statistics in a confined region. Such systems have been studied for some years from various points of view. Here we will concentrate on a class of measurements that give simultaneously the positions of all particles, as opposed to the position of a single particle. It turns out that the Pauli principle (or in other words, Fermi statistics) leads to nontrivial correlations between the particle positions. The particles are forced to group close to vertices of polygons (in two dimensional geometry) or polyhedrons (in three dimensional geometry) of unexpected shapes, named Pauli crystals.  

Geometrical structures formed by particles are common in physics -- quantum effects play a role in almost all cases. In crystalline solids quantum effects are responsible for attractive interactions between the atoms or molecules forming the crystal. In case of Wigner crystals, formed by electrons, quantum physics imposes limits on the kinetic energy of electrons linking it to the distance between them, hence to the electron density. For small densities kinetic energy is negligible compared to the electrostatic interaction energy and Wigner crystals may be observed \cite{wigner1, Wigner2}. 

In this paper we study small systems with no translational symmetry. An example from this class of systems is provided by Coulomb crystals. These structures are formed by ions in traps, see e.g. \cite{Coulomb0,Coulomb1,Coulomb2}. The structure is determined by the attractive external force of the trap and a Coulomb repulsion between the ions - quantum physics plays no role in this case. 

The difference between Pauli crystals and other similar structures lies in the role of particle interactions. Pauli crystals, as opposed to other structures, are formed by non-interacting particles. The attractive force is provided by the trap, analogously to the case of a Coulomb crystal. The exclusion principle replaces the repulsion between particles. It has been shown, however, that the Pauli exclusion principle cannot be modelled as a repulsive two body interaction \cite{2003Mullin}. This makes Pauli crystals unique, since their existence depends not on interactions but on an essential quantum effect. 

The existence of Pauli crystals was predicted in \cite{2016GajdaEPL}. Then, further discussion of these objects, routes to experimental detection, and detailed description of their properties was given \cite{2017RakshitSciRep, Symmetry}. Meanwhile, it was discussed whether at all such structures may be observed \cite{2019CiftjaAnnPhys, 2020FremlingSciRep}. Eventually Pauli crystals were successfully detected in experiments in Heidelberg \cite{PhysRevA.97.063613, PhysRevLett.126.020401}.

The present paper outlines and summarizes the methods of theoretical studies and experiments in this area. Readers are encouraged to refer to the original papers listed in the bibliography. Two geometrical scenarios are discussed -- particles in homogeneous harmonic traps in two and three dimensions and particles on the surface of a sphere.

\section{Particles in a harmonic trap}
We shall start with an example discussed already in \cite{2016GajdaEPL}. The system under consideration consists of many particles in a trap. The particles have half integer spin, hence they obey Fermi statistics. We assume that all spins are parallel, so we do not consider the spin degree of freedom. Moreover, the particles do not interact between each other. This somewhat idealized system may be realized experimentally, as discussed in a later section.

In order to make the treatment as simple as possible we describe the trap as an external harmonic potential. The system of particles is assumed to be in its ground state. Under these assumptions the state of the system is described by a many-body wavefunction in the form of a Slater determinant:
\begin{equation}
    \Psi(x_1, \cdots, x_n)=\frac{1}{\sqrt{n!}}  {\rm det}|\psi_k(x_j)|,
    \label{eq:Many_body_Psi}
\end{equation}
where $n$ denotes the number of particles, $x_i$ are particle positions, $\psi_k$ is the $k-$th energy eigenfunction in the harmonic potential. This is a well known form of the many-body wavefunction. It depends on the number of particles $n$ and also on the number of dimensions. In what follows we shall study some properties of this function.

We begin with a discussion of two particles of mass $m$ in a one dimensional harmonic trap characterized by the frequency $\omega$. The wave-function is $$\Psi(x_1,x_2)=\sqrt{\frac{1}{\pi a^{3/2}}}\left(x_1-x_2\right)\exp\left[-(x_1^2+x_2^2)/2a^2\right] $$ where $a=\sqrt{\hbar/m\omega}$.
The modulus squared of the wavefunction $|\Psi(x_1,x_2)|^2$ gives the probability density of finding the two particles at positions $x_1$ and $x_2$. It is expected that the particles will most likely be found near the maximum of this function. In this example, the two global maxima are at points $x_1=\pm a/\sqrt{2} $, $x_2=-x_1$. The maxima are indistinguishable because of the identity of the particles. Simultaneous measurement of the positions of the two particles will most probably give a result close to this maximum. Regardless of the fact that one particle is in the ground state, which has a maximum at $x=0$, it is more likely to find a particle at a position close to $x=\pm a/\sqrt{2}$ than at $x=0$. This illustrates the crucial role of quantum statistics in determining the most probable configuration.

We will now discuss particles in a two dimensional isotropic harmonic trap. In the ground state of the system the particles occupy the lowest one-particle energy states up to the Fermi level. The energy levels are degenerate and form energy shells. The ground state is not degenerate, the first excited state is doubly degenerate, the $k$-th energy level, with energy $E_k=\hbar\omega(k+1)$ contains $k+1$ one-particle states. 

\begin{figure}
\centering
  \includegraphics[width=0.5\linewidth]{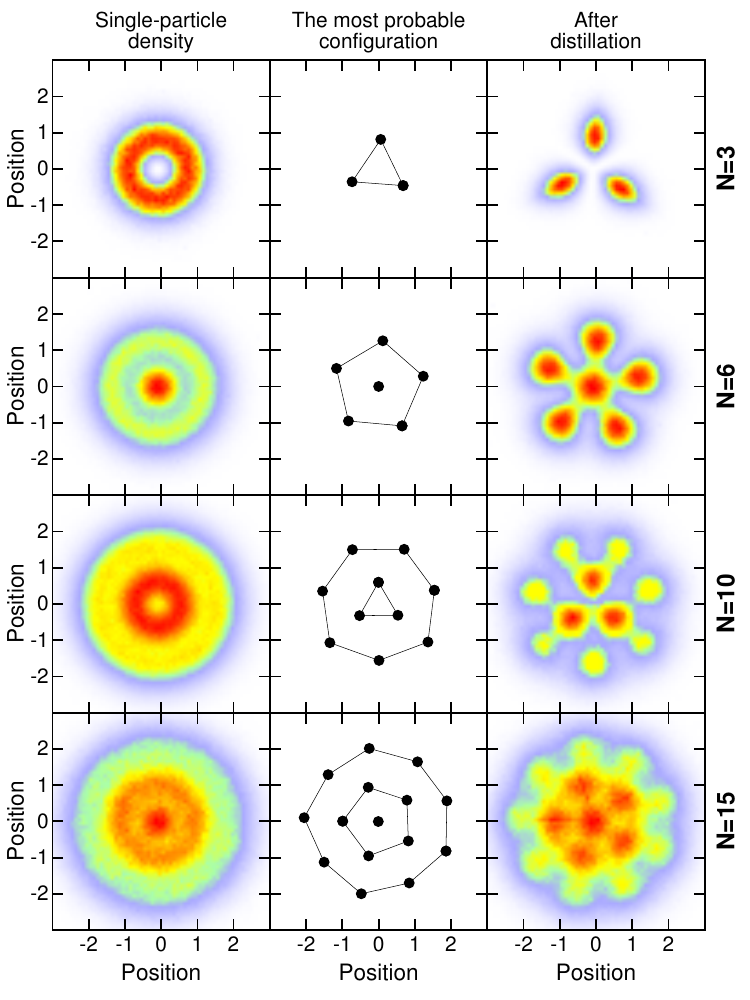}
\caption{ Pauli crystals in a two-dimensional harmonic trap. Left column: single-particle density distributions obtained from direct collection of particles’ positions in many single-shot experiments. Middle column: the most probable configuration of $n$ particles. Right column: recovered many-particle probability density. In all figures position is measured in natural units $a$ of the harmonic oscillator.}
  \label{Figure1}
\end{figure}

Let us discuss the case of three particles. One of them is in the one-particle ground state and two other particles -- in the first excited energy state. A Monte Carlo method was used to find the maximum of the modulus square of the wavefunction. It turns out that the maximum corresponds to particles at vertices of a unilateral triangle, its center coinciding with the trap center. The orientation of this triangle is arbitrary. Thus the maximum is continuously degenerate, each measurement will give positions close to one possible maximum, but its particular orientation is unpredictable. Nevertheless, at each maximum the three points $x_1, \cdots, x_3$, corresponding to positions of three particles form a unilateral triangle (Fig.~\ref{Figure1} central column).

Similar Monte Carlo calculations were used to determine the most probable configurations in case of $n=6$, $n=10$, and $n=15$ particles in a two dimensional isotropic trap. The numbers are chosen such that all energy shells are closed. In these cases the state is uniquely defined by its energy. The points giving the most probable configuration are located at vertices of polygons shown in Fig.~\ref{Figure1}. The orientation of these structures is arbitrary, it is only their shape that matters. For 6 and more particles the system develops geometrical shells. Two geometrical shells are seen in case of 6 particles. The inner shell contains one particle, while the outer shell - five particles at vertices of a regular polygon. In the case of 10 particles, two geometrical shells are present as well, with 3 and 7 particles in the inner and outer shell respectively. Note that these polygons form geometric shells which {\it are not} corresponding to the single-particle energy shells. If the number of particles is 15, three geometrical shells are observed with 1, 5 and 9 particles in consecutive geometrical shells. 

\begin{figure}
\centering
\includegraphics[width=\linewidth]{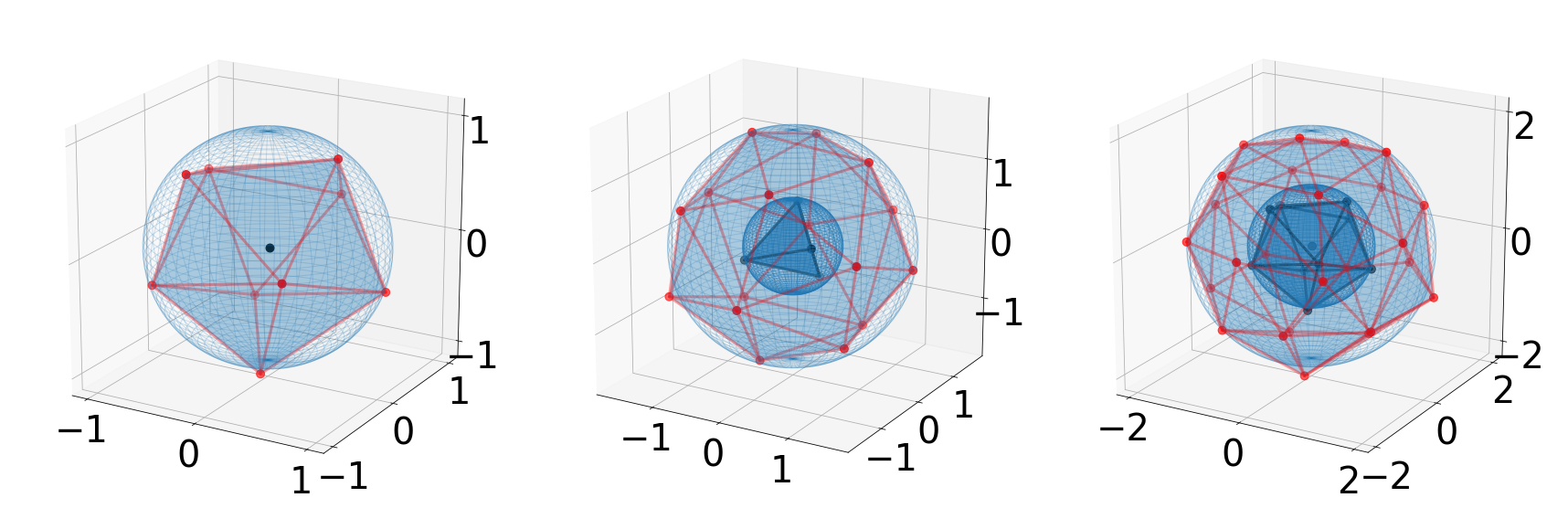}
\caption{Pauli crystals in a three-dimensional harmonic trap. The most probable configuration of $n$-particles for $n= 10, 20, 35$ (axes in natural oscillator units)} 
  \label{Figure2}
\end{figure}
A similar Monte Carlo calculation was used to find maxima of the probability distribution (wave-function modulo squared) for harmonic isotropic traps in three dimensions. The number of one particle states with energy smaller or equal to $\hbar\omega(n+3/2)$ is $(n+1)(n+2)(n+3)/6$.  As in the two-dimensional case we discuss systems with closed energy shells only ($n=1$, $4$, $10$, and $20$). The relevant structures are shown in Fig.~\ref{Figure2}. 

It is visible that the points constituting most probable configurations are located at vertices of complicated polyhedrons. Particles are grouped in geometrical shells, not related to energy shells. A ten-particle system occupies three lowest energy shells, with 1, 4 and 6 particles, but only two geometric shells, with 1 and 9 particles. A clear difference between geometrical and energy shells is also seen in the case of 20 and 35 particles. As before, the orientation of these structures in three dimensional space is arbitrary, but their shape is well defined. 

\section{Visualizing Pauli Crystals}
We will now generalize the approach described in the previous section. As mentioned above, the ground state of a system of $N$ particles is generally described in terms of a $2N$ or $3N$-dimensional complex wavefunction. Visualizing geometric properties of such a system in any form intelligible to humans will require discarding some information, and presenting the remaining data in a meaningful way. It is thus important to scrutinize what exactly is shown in such diagrams and how these figures were obtained.
As mentioned above, a very common approach is to calculate the one-particle density function, obtained from (\ref{eq:Many_body_Psi}) by integration:
\begin{equation}
    \psi_{opd}(x) = \int \Psi(x_1, \cdots, x_n) \mathrm{d}x_1 \cdots \mathrm{d}x_{n-1}
    \label{one}
\end{equation}
This function may readily be plotted, as in Fig.\ref{Figure1}, left column, but obviously it unravels very little about the many-body physics taking place.
In the second approach, we try to find the \textit{most probable configuration}. This is synonymous with finding the global maximum of (\ref{eq:Many_body_Psi}). Such plots are shown in the middle panel of Fig.\ref{Figure1}. However, these plots do not show how likely it is to find such a crystal in a single, random experiment, in other words - how sharp these maxima are. To this end, we need to sample the probability distribution resulting from (\ref{eq:Many_body_Psi}). This is done using the Metropolis-Hastings algorithm. In essence the method works as follows:
\begin{enumerate}
    \item Pick a random starting configuration $\mathbf{x} = (x_1, \cdots, x_n)$. 
    \item Calculate the probability measure of the configuration, $P = |\Psi(\mathbf{x})|^2$.
    \item Vary the configuration by a small step, $\mathbf{x'} = (x_1+\mathrm{d}x_1, \cdots, x_n +\mathrm{d}x_n)$. Calculate the new probability measure $P' = |\Psi(\mathbf{x'})|^2$.
    \item Calculate $\gamma =P'/P$ and generate a uniform random number $u$ from $[0,1]$. If $\gamma \geq u$, add $\mathbf{x'}$ to the sample and set it as the new starting point $\mathbf{x}$. Else, if $\gamma < u$, add a copy of $\mathbf{x}$ to the sample. Then return to 2. and repeat.
\end{enumerate}
Performing the random walk from step 3. can be done in a number of ways. For the case of particles on a sphere, this will be discussed in section 4.

The algorithm returns a set of many-body configurations, which can be thought of as the results of single observations ("single-shot experiments"). To visualize this data, it is not sufficient to plot a simple histogram. Because of the rotational symmetry of the Hamiltonian, the most probable configurations, as seen in Fig. \ref{Figure1}, are continuously degenerate. To unveil the underlying geometrical patterns we employ two techniques.

\paragraph{Pattern recovery}
This method was described in depth in \cite{2016GajdaEPL} - here we shall give a brief overview. Each element of the sample is "matched" to the pattern given by the most probable configuration by a sequence of isometric transformations: we first shift it by a vector $\mathbf{R_{CM}}$ so that its center of mass coincides with the center of the harmonic trap. In the next step we assign each particle of the sample element, given in polar coordinates $x_i = (r_{i}, \phi_{i})$, uniquely to a particle of the target pattern $r_{0,i} = (r_{0,\sigma(i)}, \phi_{0,\sigma(i)})$, where $\sigma$ is a suitable permutation. The distance between the two configurations is defined as

\begin{equation}
    d(\mathbf{x}, \mathbf{r_0})  = \sum^N_{i=1} (\phi_{0,i} - \phi_{\sigma(i)})^2
\end{equation}
We then rotate the configuration $\mathbf{x}$ by an angle $\alpha_0$, so that this distance is minimized. This method was applied to particles in a harmonic trap. Two-dimensional histograms of recovered patterns are presented in the third column of Fig. \ref{Figure1}. It is worth noting that this method explicitly utilizes the symmetry of the problem. 

\paragraph{Post-selection}
 This is an alternative method which does not make use of any a-priory information about the target pattern. Instead we select $n$ maxima of consecutive conditional probability distributions which eventually become vertices of a Pauli crystal. This method is used in the case discussed below, namely of particles on a sphere. As before, we begin with a sample of the many-body probability density given by the Metropolis algorithm and find a maximum of the one-particle density function, {\it i.e.} maximum of the histogram of particle positions from all configurations observed. If the histogram is a uniform function (up to shot-noise fluctuations) we can choose any point as the first vertex of the crystal. Afterwards, we select only these configurations which have a particle in the vicinity of this maximum, rejecting the other configurations. More specifically, we first find the particle of a given configuration closest to the maximum and calculate the distance $d$ between the particle position and the maximum. We then generate a random uniform number $u \in [0,1]$. The configuration is accepted into the post-selection if
\begin{equation}
    u< \frac{1}{\sqrt{2 \pi \sigma^2}} e^{-\frac{d^2}{2 \sigma^2}}
\end{equation}
Here $\sigma$ is the width of our "gaussian window" and is chosen suitably. The histogram of the post-selected configurations is a conditional probability function - the probability density of detecting another particle provided that one is located in the selected region. Due to the Pauli principle, the configurations from the post-selection will tend not to have other particles in the vicinity of previously selected maxima. In the density plot of the remaining configurations we thus see a "Pauli hole" (Fig. \ref{Figure4}, second picture). This histogram has a new global maximum which will be chosen as the position of the next vertex. If there are several global maxima, any one of them may be selected.  We apply the post-selection process to the remaining configurations and the selected maximum, and repeat as many times as we have particles. After the last iteration, the selection is a (usually small) subset of our original sample. The density plot of this selection reveals a geometrical pattern arising from high-order quantum correlations, the Pauli crystal.

This procedure will be applied in the next section to the case of particles on a sphere.


\section{Particles on a sphere}
In this section we will consider a many-body system slightly different from particles in a trap. We will assume that the particles obeying fermionic statistics and having parallel spins are bound to a sphere. The positions of these particles are given by two angles, $\theta$ and $\varphi$. In this model the single particle Hamiltonian is proportional to the square of angular momentum $H=\alpha \boldsymbol{L}^2$, where $\alpha$ is a parameter. Angular momentum is defined in the standard way: $\boldsymbol{L}=-i\hbar\left(\hat{\mathbf{e}}_\varphi\partial_\theta-\hat{\mathbf{e}}_\theta\sin^{-1}\theta\partial_\varphi\right)$, where $\hat{\mathbf{e}}_\varphi$ and $\hat{\mathbf{e}}_\theta$ denote unit vectors in the direction of $\varphi$ and $\theta$ and $\partial_\theta$ and $\partial\varphi$ denote partial derivatives with respect to the appropriate coordinate. For a particle of mass $m$ moving freely on a sphere of radius $R$, the parameter $\alpha$ is directly related to the particle's moment of inertia $1/(2\alpha)={\cal I}=m R^2$. The single-particle eigenenergies are given by $\alpha\ell(\ell+1)$ with $\ell=0,1,\ldots$. Corresponding single-particle eigenfunctions are expressed by spherical harmonics $Y_{\ell m}(\theta,\varphi)$. As in the previous examples of particles in a harmonic trap, the single-particle eigenstates are $2\ell+1$-fold degenerate. All states with the same energy $a\ell(\ell+1)$ form an energy shell labeled by the quantum number $\ell$. Closed energy shells contain $1, 3, 5,\cdots 2\ell+1$ states. The total number of states with energy smaller or equal to $\alpha\ell(\ell+1)$ is $\ell^2$. 

\begin{figure}
\centering
\includegraphics[width=\linewidth]{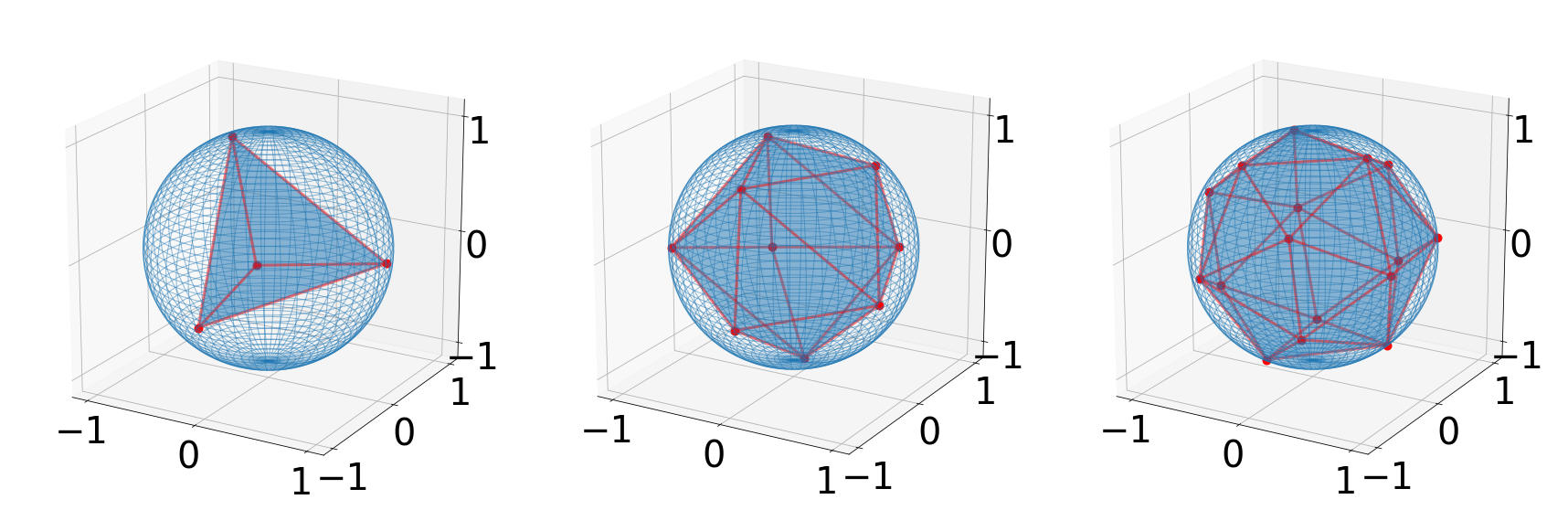}
\caption{Pauli crystals formed by fermions confined to a two-dimensional sphere. The most probable configuration of $n$-particles with $n= 4, 9, 16$.} 
  \label{Figure3}
\end{figure}

 We will discuss the case of closed shells, {\it i.e.}, when all states of energy $\alpha\ell(\ell+1)$ up to a given $\ell$ are occupied. The wave function is given by the Slater determinant of the single-particle wavefunctions. As in the case of particles in harmonic traps, the most probable configurations of particles are determined by the maxima of the modulus squared of the many-particle wavefunction. These maxima were found using a Monte Carlo method for 4, 9, and 16 particles. The most probable configurations are vertices of polyhedrons shown in Fig.~\ref{Figure3}. 
As in the previous cases, orientations of these shapes are arbitrary, the shape however is unique. It is worth noting that no trace of energy shells is present in the obtained structures.  

 To obtain configurations of maximal probability as shown in Fig. \ref{Figure3}, a random configuration of particles needs to be moved across the spherical surface. The naive approach of varying the spherical coordinates of the particles by uniformly generating random numbers $(\mathrm{d} \theta, \mathrm{d} \phi)$ would not result in a proper random walk, since it would depend on the choice of the coordinate system, i.e. the location of the north/south poles. One can see this immediately from the fact that the step size $\mathrm{d}s^2 = \mathrm{d} \theta^2 + \sin^2 \theta \mathrm{d} \phi^2$ would get smaller as the particle approached one of the poles. Therefore, a walk procedure independent of the coordinate representation is desirable. 

In our work, we use a method in which at each step, each of the particles moves in a random direction along a great circle. This is equivalent to a rotation of the point with respect to a perpendicular axis through the center of the sphere. We first generate a random, uniform point $u$ on the sphere (e.g. by setting $u_{\theta} = \arccos (2 u_1-1), u_{\phi} = 2 \pi u_2$, where $u_1, u_2 \in [0,1]$ are uniform random numbers). Given the particle position $\mathbf{r}$, we calculate the orthogonal component of $\mathbf{u}$ with respect to $\mathbf{r}$: $\mathbf{n}= \mathbf{r} - (\mathbf{r} \cdot \mathbf{u}) \mathbf{u}$. This vector, after normalization, defines the axis of rotation. We take the step length (which in our case is usually a Gaussian random number) to be the desired angle of rotation around the chosen axis.

This method of randomly moving points allows us to employ Monte Carlo procedures to find the most probable configurations shown in Fig. \ref{Figure3}, as well as generate Markovian samples as described in section 3. 

\begin{figure}
\includegraphics[width=\textwidth]{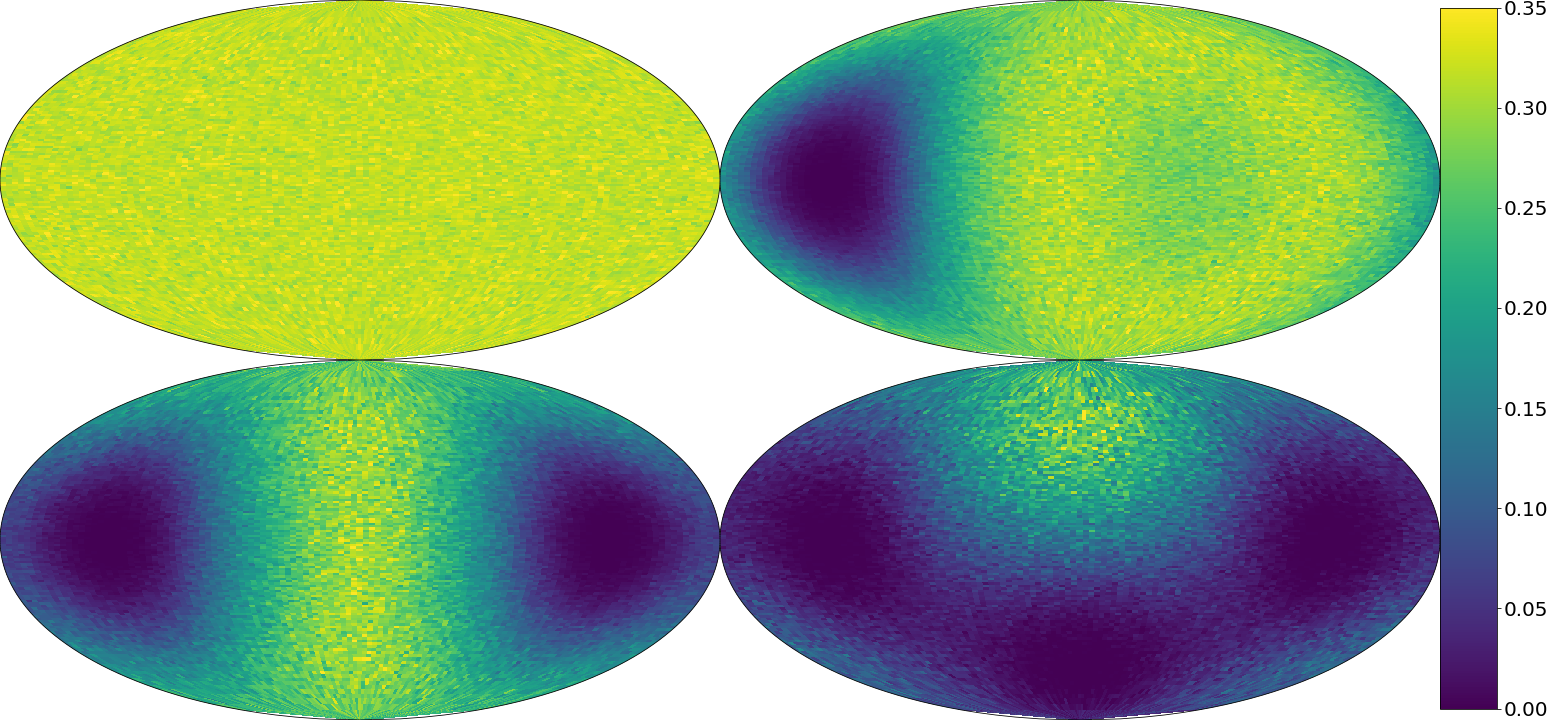}
\caption{System of four fermions confined on a two-dimensional sphere. One  particle  density function -  top left panel. The next plots represent conditional probabilities after catching 1, 2 and 3 particles in the vicinity of the maxima of the respective probability densities. Each plot is normalized to the number of remaining particles. We use the Mollweide projection of the sphere.}
\label{Figure4}
\end{figure}

\begin{figure}
\centering

\includegraphics[width=0.7\textwidth]{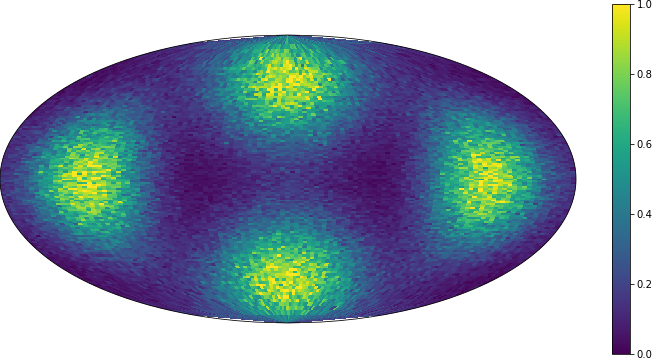}

\caption{Normalized histogram of configurations of four fermions on a sphere after all postselections are performed.}
\label{Figure5}
\end{figure}

We will employ now the postselection technique on such a sample of 4-particle-configurations on a sphere (Fig. \ref{Figure4}), which gives an alternative interpretation of the Pauli crystal. At first the distribution is uniform, as expected from the homogeneity of the spherical manifold. We can thus pick a random point as our first maximum. We choose a point on the equator of our coordinate system, since this will allow for a clear and symmetric depiction. After the first postselection the maximum of the conditional probability is now a ring around the antipode of the point selected in the first step. Due to above reasons, for the next postselection we choose a point on the ring located at the equator. After this choice, the maxima become localized, lying on a meridian through the center of our plot. We apply another postselection around one of them, so that the maximum of the final conditional probability is clearly visible. Note how in the conditional densities plots, the positions of previously selected maxima are visible as deep-blue regions of Pauli holes. Fig. \ref{Figure5} shows a histogram of 4-configurations after all postselections. The map reflects the symmetry of the Pauli crystal, which in this case is a tetrahedron. An animated, three-dimensional plot is available at \cite{sphere_gif}. Please note that such a symmetric map was obtained because of suitable choices whenever the global maximum of a conditional density, being degenerate, allowed us to choose. Regardless of such choices, the final postselected configurations would however {\it always} correspond to a regular tetrahedron, albeit rotated with respect to the distribution subjectively preferred by us for illustrative purposes.

We will end this section with a comparison of geometrical structures obtained for the three-dimensional harmonic trap with the structures on the sphere. 
It is seen in Fig.~\ref{Figure2} (showing configurations of particles in a three dimensional harmonic trap) and Fig.~\ref{Figure3} (showing configurations of particles on a sphere) that the number of particles in a given shell in a harmonic trap coincides with the number of particles required for closed energy shells in spherical geometry. Moreover, geometrical shells in a harmonic trap  correspond to spherical systems of particles filled to every second energy level. As an example let us look at the case of $10$ and $20$ particles in a harmonic trap. In the former case there is one particle in the inner shell (the number of particles required to close the first spherical energy shell), the remaining 9 particles form the outer shell (with a geometry corresponding to closing three energy shells on a sphere). In the case of $20$ particles in a trap the inner shell contains 4 particles (in a configuration given by two closed energy shells in spherical geometry) and the outer shell contains 16 particles (four closed spherical energy shells) while the shells corresponding to closed first and third energy levels are missing.

The above observation is based on a heuristic study of the emergent geometrical patterns. It may easily be checked that the postulated property is consistent with the numbers of particles forming closed energy shells on a sphere and in a three-dimensional harmonic trap. Note however that this is not a rigorous proof. 

\section{Relation between numerical modelling and experimental results}
As already mentioned, the theoretical foundations of Pauli crystals were explored in \cite{2016GajdaEPL,2017RakshitSciRep}. These papers contained also suggestions for experimental observation of these structures. Correspondingly, the experiments performed by Jochim's group in Heidelberg \cite{PhysRevLett.126.020401} dealt with particles in a two-dimensional isotropic harmonic trap. Fermionic atoms $^6{\rm Li}$ were prepared in a balanced mixture of two hyperfine states and confined by the superposition of an optical tweezer and a single layer of an optical lattice forming a two dimensional trap. In the low temperature limit all the atoms occupied the motional ground state of the harmonic trapping potential. A ramped magnetic field was used to reach the zero value of the scattering length, thus effectively switching off the interaction between the atoms. Instead of a direct measurement of atomic positions the momentum distribution was measured. The initial momenta of the particles were mapped to their positions by the time-of-flight expansion method. Positions of single atoms were detected by the fluorescence imaging scheme that allowed to detect atoms of a single-spin component with about 95\% fidelity. A single measurement gives momenta of each of the $n$ particles of given spin. 

Single measurements were thus providing many-body-configurations, geometrical structures. These measured structures were rotated as a rigid body to become as close as possible to the target pattern. In this way shapes of structures from subsequent measurements could be compared. The results clearly show the shapes predicted by the analysis of the many-body-wavefunction maxima.

\section{Conclusions}
In this paper we have presented a way of visualizing the many-body position distribution function in two or three dimensional configuration space, as opposed to $2n$ or $3n$ dimensional space. The position distribution function $|\Psi(x_1,\cdots,x_n)|^2$ determines the probability of simultaneously finding the particles at given positions. This is unlike to many experiments which focus on finding the position of single particles, determined by the one-particle distribution function.  

It is worth noting that Pauli crystals do not have the symmetry of the trap. The symmetry is broken by the process of measurement. This is a universal feature of all measurements. This symmetry breaking is particularly strong in case of open energy shells. The trap symmetry is broken not only by the measurement process but also by the state. 

Our analysis is not restricted to non-interacting particles. Similar studies may be conducted as long as the many body wave function is known. Interactions between atoms lead to complex correlations between particle positions, even though the simple correlations originating from the Pauli principle are already sufficient to exhibit the phenomenon of Pauli crystals.

\section*{Acknowledgments}
The authors thank Tomasz Sowi\'nski for suggesting the idea of Pauli crystals on a sphere and for discussions during work on the manuscript. The authors would like to thank Selim Jochim for fruitful discussions. This research was funded by the (Polish) National Science Center  Grant No. 2019/32/Z/ST2/0001 through the project  MAQS   under  QuantERA,  which  has  received funding from   the   European   Union’s Horizon   2020 research and innovation program under grant agreement no 731473 (FG and MG).

\end{document}